# MEASUREMENT OF SOFTWARE DEVELOPMENT EFFORT ESTIMATION BIAS: AVOIDING BIASED MEASURES OF ESTIMATION BIAS

Magne Jørgensen

Simula Metropolitan Center for Digital Engineering, Oslo, Norway
magnej@simula.no

## *ABSTRACT*

*In this paper, we propose improvements in how estimation bias, e.g., the tendency towards under-estimating the effort, is measured. The proposed approach emphasizes the need to know what the estimates are meant to represent, i.e., the type of estimate we evaluate and the need for a match between the type of estimate given and the bias measure used. We show that even perfect estimates of the mean effort will not lead to an expectation of zero estimation bias when applying the frequently used bias measure: (actual effort – estimated effort)/actual effort. This measure will instead reward under-estimates of the mean effort. We also provide examples of bias measures that match estimates of the mean and the median effort, and argue that there are, in general, no practical bias measures for estimates of the most likely effort. The paper concludes with implications for the evaluation of bias of software development effort estimates.*

## *KEYWORDS*

*Software development effort estimation, measurement of estimation overrun, proper measurement of bias*

## 1. INTRODUCTION

Imagine that you are asked to throw two dice many times. Each time, the outcomes of the two dice are multiplied. The dice are fair, which gives 36 equally likely outcomes ranging from 1 (1x1) to 36 (6x6), with a mean outcome of 12.25 (3.5x3.5), a median outcome of 10 (half of the outcomes have values of 10 or less) and a mode of 6 (most frequently occurring value[1]). What would be the best estimates of the products when the goal is to give unbiased estimates?

A good response would be that the best estimates depend on how the bias will be measured, i.e., how we will interpret unbiased estimates. We may, for example, interpret estimation bias as mean bias (a tendency to give estimates higher or lower than the mean), median bias (a tendency to give estimates higher or lower than the median), or even mode bias (a tendency to give estimates higher or lower than most frequently occurring value). Assume that you get informed that the goal is to give mean unbiased estimates, and consequently select the mean outcome (12.25) for each of your estimates. You are likely to have the correct prediction less often compared to the use of the mode (6) as your estimate and more than half of your estimates will be too high, but the mean outcome will be expected to be the same as your estimate, i.e., the bias measure:

$$\frac{1}{N}\sum_{i=1}^{N}(act_i - est_i) = \frac{1}{N}\sum_{i=1}^{N}(act_i) - \frac{1}{N}\sum_{i=1}^{N}(est_i)$$

has an expected value of zero, where $act_i$ and $est_i$ are the actual and estimated product of throw $i$, respectively.

---

[1] The value 12 is occurring equally often, but for simplicity, we choose 6 as our mode value in this example.

What if you had used the relative (scale-free) measure:

$$\frac{1}{N}\sum_{i=1}^{N} \frac{(act_i - est_i)}{act_i}$$

as your bias evaluation measure. In this case, as we will show later, neither the mean nor the median gives an expected bias of zero. Instead, this measure rewards under-estimating the mean a lot. The optimal value for this bias measure is in our example is 6, which happens to be the mode (but will not in general be the mode). If we, however, change the bias measure to:

$$\frac{1}{N}\sum_{i=1}^{N} \frac{(act_i - est_i)}{est_i}$$

i.e., divide by the estimated rather than the actual values, the mean again becomes the estimate leading to an expectation of zero bias. The use of the median outcome as your estimate is for none of the three bias measures leading to an expectation of zero bias.

The above example is meant to demonstrate that without a match between type of estimates and how we evaluate the bias of the estimates, we cannot ensure evaluation fairness (giving perfect scores to perfect estimates of the intended type) and we risk rewarding poorer estimation performance. Implicitly, the example also tells us that without knowing what type of estimate we evaluate, it is hard to know how to evaluate it properly.

The example can easily be transformed into one of effort estimation. It illustrates, for example, that even perfect estimates of the mean effort will not give an expected estimation bias of zero when using the measure:

$$\frac{1}{N}\sum_{i=1}^{N} \frac{(act_i - est_i)}{act_i}$$

and that observing an estimation bias of zero using this measure suggests that the estimates had a tendency to under-estimate the mean effort, not that the estimates are unbiased. Not knowing what type of effort estimate we evaluate makes it hard to evaluate their bias, e.g., degree of under-estimating the actual effort, properly.

The next section aims at giving more insight into the match and lack of match between the type of estimates and frequently used bias measures. While there has been much research on challenges with the use of common software development estimation *error* measures, in particular the challenge connected with the use of the Mean Magnitude of Relative Error (MMRE), see for example [1-5], we have been unable find analyses of challenges and proper use of software development effort estimation *bias* measures.

## 2. MEASURES OF THE BIAS OF EFFORT ESTIMATES

We propose that to enable fair evaluation of the bias, such as the degree of overrun, of effort estimates we need to:

- Identify the type of estimates to be evaluated, e.g., whether the estimates are intended to be estimates of the mean, median or mode (most likely) use of effort.
- Select a bias evaluation measure that matches the identified type of estimate. A match is in this context understood as that the measure gives zero bias for perfect estimates of the identified type of estimate. This criterion corresponds to what is the requirement for a proper scoring rule, see for example [6].

These conditions for meaningful evaluation of estimation bias are similar to those suggested for evaluation of estimation error, in [7, 8], i.e., the same conditions applied on bias instead of error.

Notice that these requirements are not meant to be sufficient, just necessary conditions, for meaningful evaluation of estimation bias.

For the first step, identifying the type of estimate, we believe it is useful, perhaps even necessary, to apply a probabilistic view on effort usage, i.e., that the use of effort is uncertain and that the actual effort is one sample (draw) from a (typically unknown) probability effort distribution. Common types of estimates, with probabilistic interpretations, are estimates of the mean, the median, and the most likely (mode) effort. Estimates of the mean effort is, for example, what is intended produced by linear regression estimation models derived by using OLS (Optimized Linear Square), the median effort is frequently used as the planned effort or as input to the budget in several large-scale projects [9], and the most likely effort may be what the software developers give when they are asked about software task estimate [10]. The widely used PERT-model, see [11], is a good example of the use of different types of estimates with probabilistic interpretations in software development effort estimation contexts. Here the developers are asked to give the mode[2] (the most likely effort), together with the minimum and the maximum effort (or the 10 and 90% percentiles). The mean effort is then calculated based on these three values, typically using the formula:

$$mean\ effort = \frac{min.\ effort + 4\ most\ likely\ effort\ + max.\ effort}{6}$$

and used in the planning of the software projects [11].

In the following, we briefly discuss a selection of non-matching and matching bias measures for each of the three types of estimates, i.e., estimates of the mean, median and mode.

## 2.1. Assessment of bias of estimates of the mean effort

We start by showing, as claimed in the introduction, that the commonly used estimation bias (effort overrun) measure:

$$mean\ RE_{act} = \frac{1}{N} \sum_{i=1}^{N} \frac{(act_i - est_i)}{act_i}$$

rewards underestimates of the mean efforts.

Assume that we use the true mean effort as our estimate, i.e., we set the estimated effort ($est_i$) equal to the mean value ($\mu_i$) of the software development effort distributions $i = 1…N$. In other words, we have perfect estimates of the mean efforts. We then have that the expected value of this measure, due to the linearity of the mean, can be expressed as:

$$E\left[\frac{1}{N}\sum_{i=1}^{N}\frac{(act_i - \mu_i)}{act_i}\right] = \frac{1}{N} \cdot \left[E\left(\frac{act_1 - \mu_1}{act_1}\right) + \cdots + E\left(\frac{act_N - \mu_N}{act_N}\right)\right] =$$

$$\frac{1}{N} \cdot \left[E\left(1 - \frac{\mu_1}{act_1}\right) + \cdots + E\left(1 - \frac{\mu_N}{act_N}\right)\right] = 1 - \frac{1}{N} \cdot \left[E\left(\frac{\mu_1}{act_1}\right) + \cdots + E\left(\frac{\mu_N}{act_N}\right)\right].$$

An approximation of the expected value of a ratio of stochastic variables, see for example [12], is:

---

[2] The reason for not requesting the mean estimate, but instead the mode, may be that it is believed to be easier to give qualified expert judgments on the effort typically needed for the type of task (the most likely or mode effort), rather than the (perhaps more abstract in terms of looking back on prior experience) estimate of the mean use of effort.

$$E\left(\frac{X}{Y}\right) \approx \frac{\mu_X}{\mu_Y} - \frac{Cov(X,Y)}{\mu_Y^2} + \frac{Var(Y)\mu_X}{\mu_Y^3}$$

Replacing X with the mean ($\mu_i$), which is our estimate of the *i*-th task, and Y with the random variable *act$_i$* (which by definition has as its expected (mean) value $\mu_i$) give that, for all tasks *i*:

$$\left(1 - E\left(\frac{\mu_i}{act_i}\right)\right) \approx 1 - \frac{\mu_i}{\mu_i} - \frac{Cov(\mu_i, act_i)}{\mu_i} + \frac{Var(act_i)\mu_i}{\mu_i^3} = \frac{Cov(\mu_i, act_i)}{\mu_i} + \frac{Var(act_i)}{\mu_i^2}$$

We have that $Cov(\mu_i, act_i) = 0$, since the correlation between the mean ($\mu_i$) and the sampled values (*act$_i$*) will be zero. This implies that we should, for a perfect estimate of the mean effort, expect a bias towards over-estimation of effort of size:

$$\frac{Var(act_i)}{\mu_i^2}$$

when using the mean RE$_{act}$ as our measure of bias, i.e., the bias of the bias measure (when evaluating estimates of the mean effort) increases with the variance of the actual effort.

Interestingly, if we use the slightly modified measure:

$$mean\ RE_{est} = \frac{1}{N}\sum_{i=1}^{N} \frac{(act_i - est_i)}{est_i}$$

i.e., when we divide by the estimated rather than the actual effort, as implemented in for example [13], we have an expected bias of zero when the mean efforts are used as the estimates:

$$E\left[\frac{1}{N}\sum_{i=1}^{N} \frac{(act_i - \mu_i)}{\mu_i}\right] = \frac{1}{N} \cdot \left[E\left(\frac{act_1 - \mu_1}{\mu_1}\right) + \cdots + E\left(\frac{act_N - \mu_N}{\mu_N}\right)\right]$$
$$= \frac{1}{N} \cdot \left[E(act_1 - \mu_1) \cdot E\left(\frac{1}{\mu_1}\right) + \cdots + E(act_N - \mu_N) \cdot E\left(\frac{1}{\mu_N}\right)\right]$$
$$= \frac{1}{N} \cdot \left[(E(\mu_1) - E(\mu_1)) \cdot E\left(\frac{1}{\mu_1}\right) + \cdots + (E(\mu_N) - E(\mu_N)) \cdot E\left(\frac{1}{\mu_N}\right)\right] = 0$$

due to independence between $(act_i - \mu_i)$ and $\mu_i$.

Perfect estimates of the mean effort also, due to the linearity of the mean values and as exemplified in the introduction, give zero bias for:

$$mean\ RE_{dev} = \frac{1}{N}\sum_{i=1}^{N}(act_i - est_i).$$

Our results consequently shows that we should use:

$$\frac{1}{N}\sum_{i=1}^{N} \frac{(act_i - est_i)}{est_i} \text{ or } \frac{1}{N}\sum_{i=1}^{N}(act_i - est_i)$$

as our estimation bias measure if we want to give zero bias to perfect estimates of the mean. It also shows that we should stop using

$$\frac{1}{N}\sum_{i=1}^{N} \frac{(act_i - est_i)}{act_i}$$

as a bias measure when evaluating estimates of the mean. Using this bias measure, we will not only evaluate perfect estimates of the mean as biased towards over-estimation, but also reward under-estimation of the mean effort. For example, a seemingly unbiased estimation model, giving a mean RE$_{est}$ of zero, has in reality given estimates with a bias towards under-estimating the mean effort with the value of:

$$\frac{Var(act_i)}{\mu_i^2}$$

per estimate.

## 2.2 Assessment of bias of estimates of the median effort

Measurement of the bias of estimates of the median effort, with the requirement of zero bias for perfect estimates of the median, is trivial for bias measures based on the median, rather than the mean, deviation.

In such cases, we have that the median of all three measures:

$$(act_i - est_i), \frac{(act_i - est_i)}{act_i} \text{ or } \frac{(act_i - est_i)}{est_i}, i = 1..N,$$

results in zero bias for perfect estimates of the median effort. This is the case since the median, by definition, is the value that is just as likely to exceed as not to exceed, i.e., half of the observations will have positive and the other half negative deviations between the actual and the estimated values. Dividing this deviation by the estimated or the actual value does not affect this relationship. All the above measures are consequently proper measures of median bias of effort estimates.

Another median-matching bias measure is based on the logarithm of the ratio of the estimated and the actual effort (log-error). This measure, proposed in amongst others [14], has the advantage that it combines being a relative (scale-free) measure with symmetric bias values. As is easy to see, the relative bias measures:

$$\frac{(act_i - est_i)}{act_i} \text{ and } \frac{(act_i - est_i)}{est_i}$$

are *not* symmetric around 0, i.e., the possible scores are in the interval (-∞, 1) and (-1, ∞), respectively, which for large estimation errors give much higher penalties for deviations in one direction compared to the other.

The proposed, symmetric and median unbiased, measure, which we term MdLogErr, is defined as the:

$$median \text{ } of \log\left(\frac{act_i}{est_i}\right) = median \text{ } of \text{ } \log(act_i) - \log(est_i), for \text{ } i = 1 \dots N,$$

The expected value of MdLogErr is zero for perfect estimates of the median due to the preservation of the percentiles, including the median (which is the 50% percentile) when back-transforming the log-transformed values, i.e., if 50% of *log (act$_i$)* is above *log(median$_i$)*, then 50% of the *act$_i$* will also be above the *median$_i$*. Use of the log-error may be said to have non-intuitive interpretation of bias, given that it is based on log-scores rather than percentage deviation. The interpretation of the scores of asymmetric bias measures is, however, not trivial either, given the different penalties for high over- and under-run. A possible selection criterion is to use asymmetric measures like

$$\frac{(act_i - est_i)}{est_i}$$

when there are no large deviations between the estimated and actual effort expected, in which case the asymmetry would not complicate the interpretations, and to select

$$log\left(\frac{act_i}{est_i}\right)$$

otherwise.

## 2.3 Assessment of bias of estimates of the mode effort

The assessment of bias of estimates of the mode (most likely) effort introduces several evaluation challenges. If we know the percentiles of the modes, we may (similarly to how we evaluate the median) measure mode bias through measures of calibration and informativeness (or sharpness), see for example [7, 15]. For example, if the mode values, on average, were at the 45% percentile, then unbiased estimates of the mode should be higher than exactly 45% of the actual effort values. Unfortunately, we typically do not know the percentiles of the mode values.

As pointed out in several studies, see for example [16, 17], there are inherent problems in evaluating predictions of the mode. To our knowledge, there are currently no practical measures enabling bias evaluations of estimates of the mode, other than through the hit rate of its percentiles.

## 3. IMPLICATIONS AND CONCLUSION

Possibly, the great majority of previous studies on software development effort estimation bias may be accused of doing what Gneiting [6] describes as: "*The common practice of requesting 'some' point forecast, and then evaluating the forecasters by using 'some' (set of) scoring function(s), is not a meaningful endeavor.*" As demonstrated in this paper, meaningful evaluation of estimation bias requires both that we know what we evaluate and that we select bias evaluation measures that match the type of estimates we evaluate, i.e., that we have a proper scoring rule. Otherwise, we may produce misleading results and/or give incentives for strategically too low or too high estimates. If, for example, an estimator knows that the estimates will be evaluated by its average percentage overrun, i.e., the measure:

$$mean\ RE_{act} = \frac{1}{N}\sum_{i=1}^{N}\frac{(act_i - est_i)}{act_i}$$

the estimator will benefit from giving estimates lower than the mean effort instead of giving honest estimates of the mean. The higher the variance in the use of effort per project or task, the lower are the estimates that give an expectation of zero average effort overrun using mean $RE_{act}$.

As an illustration of the possible implications of different types of estimates on measures of bias, assume that the actual use of effort of a hypothetical software project is distributed as in Figure 1. The distribution in Figure 1 is a log-normal distribution[3] with mean of 236 work-hours, a median of 209 work-hours, a mode (most likely effort) of 162 work-hours, and a standard deviation of 126 work-hours. We calculated the expected estimation bias by simulating that the project is executed 10.000 times and that the actual effort is for each project execution randomly drawn from the distribution in Figure 1, i.e., we assume a hypothetical repeated execution of the same project without learning in the same context. We then calculated the expected (mean) $RE_{act}$ in the case of that we use perfect estimates[4] of the mode, the median or the mean of the actual effort distribution as our effort estimate. From this simulation we found that if the estimator gave the most likely effort (162 work-hours) as his/her project effort estimate, the expected $RE_{act}$ would be 12% (12% effort overrun), while the use of the median effort (209 work-hours) as the estimate would give an expected $RE_{act}$ of -14% (14% effort underrun), and use of the mean effort (236 work-hours) as the estimate would give an expected $RE_{act}$ of -28% (28% effort underrun). The 28% effort underrun is very close to what is expected from the formula we derived earlier, i.e., the expected effort underrun when estimating the mean effort and using the $RE_{act}$ as our measure of estimation bias is:

$$\frac{Var(act_i)}{\mu_i^2} = \frac{126^2}{236^2} = 0.29$$

---

[3] As argued in, for example [10], a log-normal distribution reflects typical properties of effort usage distributions, e.g., that it is right-skewed and has a minimum of zero.

[4] Clearly, in realistic situations we do not know with high accuracy the distribution of actual effort of a software project, but have to estimate this, as well. For illustrative purposes on properties of the bias measure, however, the assumption of perfect knowledge about the underlying effort distribution (effort uncertainty) is considered useful. Similarly, while we will never repeat the same project without learning in the same context, this assumption is useful to illustrate the properties of the bias measures. The core result do not change if we, for example, replace the actual mean of the effort distribution (assuming a perfect estimator) with the estimated mean (assuming that the estimator just have an estimate of the underlying mean use of effort), or that we replace the assumption of repeating the same project without learning with executing different projects and learning.

The estimate that gives an expected $RE_{act}$ of zero is in this case approximately 185 work-hours, which is neither the mode, the median or the mean of the distribution of actual effort.

Our illustration documents that the measured bias varies much dependent on what type of estimate the developer gives. Also, the illustration shows that it will be hard to know what type of estimate to give, to have an expectation of zero bias, when using the bias measure $RE_{act}$. Giving perfect estimates of the mean effort of a project will, for example, always[5] result in an expected $RE_{act}$ larger than zero (effort underrun).

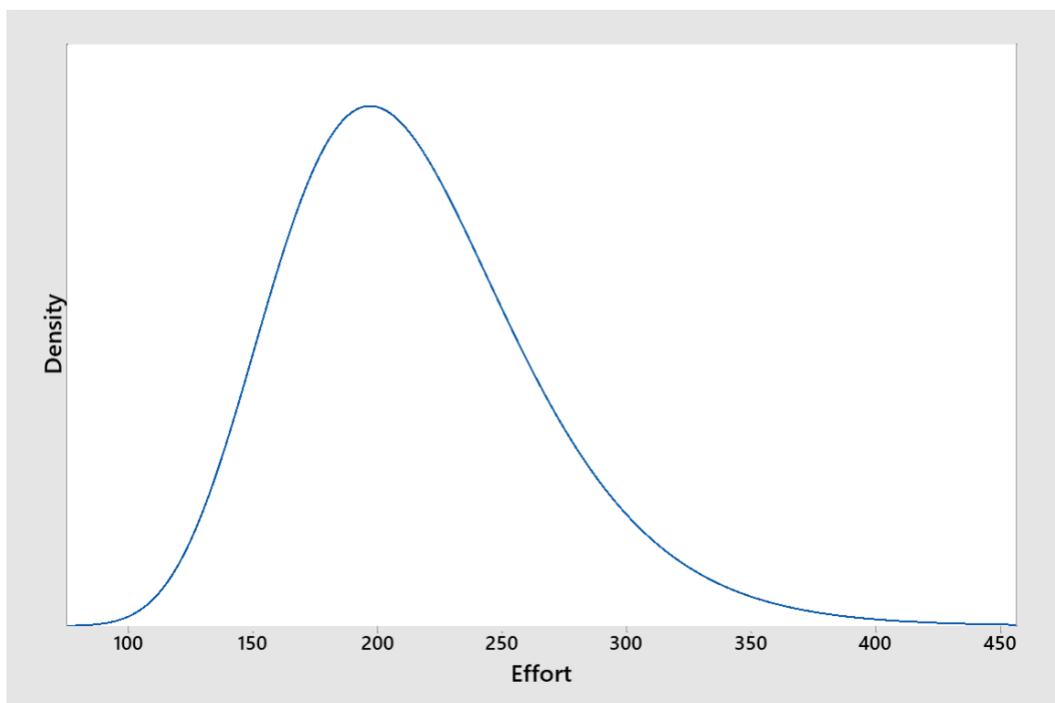

Figure 1. Hypothetical distribution of effort usage

In many estimation bias evaluation contexts, the type of estimates to be evaluated will not be explicitly described, but instead simply be presented as effort 'estimates'. In those contexts, to enable fair and meaningful bias evaluations, we will have to make qualified guesses about what the estimates are meant to represent. Such qualified guesses may be made derived from the optimization function used to derive the model (e.g., OLS-regression, PERT-formula, etc.), the purpose of the estimate (e.g., input to budget or plan), or the estimation instructions (e.g., requesting the "most likely" use of effort). If the estimates to be evaluated are believed to be of various types we may either evaluate different types of estimates separately or evaluate the estimates based on the dominant type of estimate.

---

[5] It is only when there is no variance in the use of effort that perfect estimates of the mean will give zero bias. Clearly, no variance in use of effort is unrealistic in the context of software development projects.

To continue as before, with little or no reflection on the type of estimate or concerns about a match between the type of estimate and evaluation measure, means that we will continue having a hard time interpreting how biased the effort estimates are and how large the cost overruns in software development really are.

## REFERENCES

[1] J. S. Armstrong, "A commentary on error measures," *International Journal of Forecasting,* vol. 8, pp. 99-111, 1992.

[2] T. Foss, E. Stensrud, B. Kitchenham, and I. Myrtveit, "A simulation study of the model evaluation criterion MMRE," *Ieee Transactions on Software Engineering,* vol. 29, no. 11, pp. 985-995, 2003.

[3] B. A. Kitchenham, L. M. Pickard, S. G. MacDonell, and M. J. Shepperd, "What accuracy statistics really measure," *IEE Proceedings-Software,* vol. 148, no. 3, pp. 81-85, 2001.

[4] I. Myrtveit and E. Stensrud, "A controlled experiment to assess the benefits of estimating with analogy and regression models," *Ieee Transactions on Software Engineering,* vol. 25, no. 4, pp. 510-525, 1999.

[5] M. Jørgensen, "A critique of how we measure and interpret the accuracy of software development effort estimation," in *First international workshop on software productivity analysis and cost estimation*, 2007: Citeseer.

[6] T. Gneiting, "Making and evaluating point forecasts," *Journal of the American Statistical Association,* vol. 106, no. 494, pp. 746-762, 2011.

[7] M. Jørgensen, "Evaluating probabilistic software development effort estimates: Maximizing informativeness subject to calibration," *Information and software Technology,* vol. 115, pp. 93-96, 2019.

[8] M. Jørgensen, M. Welde, and T. Halkjelsvik, "Evaluation of Probabilistic Project Cost Estimates," *IEEE Transactions on Engineering Management,* vol. In press, 2021.

[9] K. F. Samset, G. H. Volden, N. Olsson, and E. V. Kvalheim, "Governance schemes for major public investment projects: A comparative study of principles and practices in six countries," ed: Ex ante akademisk forlag, 2016.

[10] T. Halkjelsvik and M. Jørgensen, *Time Predictions: Understanding and Avoiding Unrealism in Project Planning and Everyday Life*. Springer, 2018.

[11] D. Golenko-Ginzburg, "On the distribution of activity time in PERT," *Journal of the Operational Research Society,* vol. 39, no. 8, pp. 767-771, 1988.

[12] A. Stuart, S. Arnold, J. K. Ord, A. O'Hagan, and J. Forster, *Kendall's advanced theory of statistics*. Wiley, 1994.

[13] K.-S. Na, J. T. Simpson, X. Li, T. Singh, and K.-Y. Kim, "Software development risk and project performance measurement: Evidence in Korea," *Journal of Systems and Software,* vol. 80, no. 4, pp. 596-605, 2007.

[14] L. Törnqvist, P. Vartia, and Y. O. Vartia, "How should relative changes be measured?," *The American Statistician,* vol. 39, no. 1, pp. 43-46, 1985.

[15] T. Gneiting and M. Katzfuss, "Probabilistic forecasting," *Annual Review of Statistics and Its Application,* vol. 1, pp. 125-151, 2014.

[16] T. Gneiting, "When is the mode functional the Bayes classifier?," *Stat,* vol. 6, no. 1, pp. 204-206, 2017.

[17] C. Heinrich, "The mode functional is not elicitable," *Biometrika,* vol. 101, no. 1, pp. 245-251, 2014.

**Author**

Magne Jørgensen received a Dr. Scient degree in informatics from University of Oslo in 1994. He has 10 years of industry experience as software developer, project leader and manager and worked for 20 years as a professor at University of Oslo. He is currently a chief research scientist at Simula Metropolitan and a professor at Oslo Metropolitan University. He is one of the founders of evidence-based software engineering. Current research interests include software management, psychology of human judgement and cost estimation. (H-index 52).

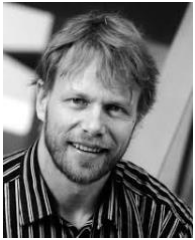